\documentclass{evn2002}
\usepackage{graphicx}
\begin{document}
   \title{VLBI observations at 147 GHz: first detection of transatlantic fringes in bright AGN}

   \author{
           T.P. Krichbaum, D.A. Graham,  W. Alef, A. Polatidis, U. Bach, A. Witzel, J.A. Zensus\inst{1} 
           \and \\
           A. Greve, M. Grewing\inst{2} 
           \and \\
           S. Doeleman, R. Phillips, A.E.E. Rogers, M. Titus\inst{3}
           \and \\
           H. Fagg, P. Strittmatter, T.L. Wilson, L. Ziurys\inst{4}
           \and \\
           R. Freund\inst{5}
           \and \\
           P. K\"on\"onen, J. Peltonen, S. Urpo\inst{6}
           \and \\
           F. Rantakyro\inst{7,8}, J. Conway\inst{8} 
           \and 
           R.S. Booth\inst{8}
          }

   \institute{
           Max-Planck-Institut f\"ur Radioastronomie, Auf dem H\"ugel 69, 53121 Bonn, Germany
           \and
           Institut de Radioastronomie Millim\'etrique, 300 Rue de la Piscine, 38460 St. Martin d'H\`eres, Grenoble, France
           \and
           Massachusetts Institute of Technology, Haystack Observatory, Off Route 40, Westford, MA 01886, USA
           \and
           Steward Observatory, University of Arizona, Tucson, AZ 85721, USA
           \and
           National Radio Astronomy Observatory, 949 North Cherry Avenue, Tucson, AZ 85721, USA
           \and
           Mets\"ahovi Radio Research Station, Helsinki University of Technology, 02540 Kylm\"al\"a, Finland
           \and
           Swedish-ESO Submillimetre Telescope, ESO, St. Alonso de Cordova 3107, Vitacura, Casilla 19001, Santiago 19, Chile.
           \and
           Onsala Space Observatory, 43992 Onsala, Sweden
             }

   \abstract{
   At 147\,GHz (2\,mm wavelength), we detected three prominent AGN (NRAO\,150, 3C\,279, 1633+382)
   with {\bf V}ery {\bf L}ong {\bf B}aseline {\bf I}nterferometry (VLBI) with
   an angular resolution of only $\sim 18$ micro-arcseconds. This is a new world record
   in radio interferometry and astronomical imaging and opens fascinating future possibilities to directly
   image and study the innermost regions in Quasars and other Active Galactic Nuclei.
   }

\titlerunning{VLBI at 147 GHz}
\authorrunning{Krichbaum et al.}

   \maketitle
%
%________________________________________________________________

\section{Introduction}
Even after more than 40 years after Marten Schmidt's discovery
of the cosmological redshift of the hydrogen lines in 3C\,273, and of comprehensive astrophysical 
research on Active Galactic Nuclei (AGN), 
the enigma of the origin of their extreme luminosity (ranging from radio to
Gamma-ray bands) and the creation mechanism for the highly relativistic plasma jets 
(often extending over many hundred kpc) is still not solved. Although the majority
of the scientific community regards accretion onto supermassive Black Holes as the
most plausible explanation for the `quasar phenomenon', 
many details of the astrophysical processes taking place in the centers of these most luminous 
objects in the Universe still remain unexplained.
In particular is the question of how the relativistic jets are made, accelerated and confined
not satisfyingly answered. In order to test existing theories,
most of which try to explain energy release and jet production
via coupling to the accretion process onto a supermassive Black Hole
(e.g. the Blandford-Payne magnetic sling-shot mechanism), the direct imaging of the innermost 
regions of AGN becomes of great importance. The technique of 
interferometry is the only astronomical observing method, which leads to such direct images.

In {\bf V}ery {\bf L}ong {\bf B}aseline {\bf I}nterferometry (VLBI) the angular resolution
can be improved, either by increasing the distance between the radio telescopes or
by observing at shorter wavelengths. The first approach leads to VLBI with orbiting radio antennas 
in space (e.g. VSOP, Hirabayashi et al. 2000), which however at present gives only an angular resolution
of 0.2--0.3\,mas (1\,mas = $10^{-3}$ arcsec) at 5\,GHz. 
The second possibility leads to VLBI at millimeter wavelengths (mm-VLBI),
which furthermore facilitates the imaging of compact structures, which are 
self-absorbed (opaque), and therefore not directly observable, at the longer centimeter wavelengths.

Nowadays, mm-VLBI observations are regularly performed at 86\,GHz ($\lambda=3.5$\,mm), where
images with angular resolutions of up to $\sim 50 \mu$as ($1 \mu$as = $10^{-6}$ arcsec) are obtained
(e.g. Rantakyro et al. 1998, Lobanov et al. 2000).\\
VLBI observations at even shorter wavelengths are technically more difficult and have not yet
passed the stage of test experiments on relatively short continental baselines. In 1989 and at the so far
highest VLBI frequency of 223\,GHz, the quasar 3C\,273 was marginally detected (with SNR $\leq 7$) on the 
baselines Owens Valley to Kitt Peak (845 km, 0.65G$\lambda$) (Padin et al. 1990). In 1994 and at 215\,GHz,
first fringes were seen between the IRAM 30\,m antenna on Pico Veleta (Spain) and a single antenna
of the IRAM interferometer on Plateau de Bure (France) (Greve et al. 1995). A second
experiment on this 1147 km (0.88G$\lambda$) long baseline in 1995, 
resulted in the successful (SNR $\leq 35$) VLBI detection 
of 8 out of 9 observed compact flat spectrum sources, including the source in the  Galactic Center Sgr\,A* 
(Krichbaum et al. 1997). This experiment lead to a number of conclusions, which are important for the future: 
(i) a large fraction of the known cm-VLBI sources are compact enough, so that they can be observed 
with VLBI at short millimeter wavelengths, 
(ii) the VLBI jets can be traced to sub-parsec scales, however, the curvature of the jets usually increases
towards the nucleus, and (iii) the internal structure of the Galactic Center source Sgr\,A* becomes visible 
through the foreground scattering IGM, and the size of Sgr\,A* must be smaller than 
$\sim  20$ Schwarzschild radii (Krichbaum et al. 1998). 

Between 1995 -- 2000 several attempts with various telescopes were made to achieve VLBI detections 
on the longer transatlantic baselines. These experiments were performed in the 2\,mm and 1.3\,mm bands,
but failed due to technical difficulties. The recent promising detection of 3C\,273 and 3C\,279 at 147\,GHz
on the 3100 km (1.5G$\lambda$) baseline between Pico Veleta and Mets\"ahovi (Finland) in March/April 2001 
(Greve et al. 2002), and the availability of VLBI equipment and a new 2\,mm receiver at the Heinrich
Hertz telescope (HHT) on Mt. Graham (Arizona), stimulated a transatlantic VLBI experiment at 147\,GHz, 
which we will describe in the following.

\section{The Experiment, Data Analysis, Transatlantic Fringe Detection}

The VLBI experiment was performed at 147 GHz and in coordination with a spectral line VLBI experiment
at 129 GHz (see Doeleman et al. , this conference). The 147 GHz observations were done on April 18, 15 UT  - April 19, 6 UT. 
Participating telescopes were the 30\,m IRAM antenna on Pico Veleta (PV) in Spain, the 10\,m Heinrich-Hertz
telescope (HHT) on Mt. Graham (Arizona),  the 12\,m telescope on Kitt Peak (KP) (Arizona), the 14\,m antenna at
Mets\"ahovi (MET) (Finland) and the 15\,m SEST telescope in La Silla  (Chile). Table 1 summarizes
the antenna properties. 

\begin{table}
\begin{tabular}{l|c|c|c|c|r}
Telescope      &  h   & D  &${\rm T_{sys}}$ &$\eta$& SEFD  \\
               &  [m] &[m] &  [K]           &      &  [Jy] \\ \hline
Pico Veleta    & 2900 & 30 &~~200           & 0.50 &  1600 \\
Heinrich-Hertz & 3200 & 10 &~~200           & 0.60 & 11700 \\
Kitt Peak      & 1900 & 12 &~~250           & 0.45 & 13600 \\
Mets\"ahovi    &  40  & 14 &$\sim$1000      & 0.13 &140000 \\
SEST           & 2300 & 15 &~~350           & 0.45 & 12200 \\ 
\end{tabular}
\caption{Antenna properties: tabulated are the name (col.1), the altitude (col.2), the antenna diameter
(col. 3), and the typical system temperature (col. 4), aperture efficiency (col.5) and
system equivalent flux density at 147 GHz.}
\vspace{-0.5cm}
\end{table}

The observations were performed at a reference frequency of 147028.99 MHz. The data were recorded with the 
MKIV VLBI system at 224 Mbit/sec in 2 bit mode (256-8-2). Due to the limited  number of only 4 video 
converters at HHT and KP, the frequency synthesis was made by recording upper and lower side band
in each of the four base band converters (BBC's). The total observing bandwidth was 56\,MHz, since the 
lower sideband in the first BBC was not recorded. To produce left circular
polarization (LCP) quarter-wave plates were inserted in front of each receiver (single sideband tuned
4\,K cooled SIS systems). Special care was taken
for the tuning and stability of the LO-systems. Test tones were injected and round-trip phase stability tests
were performed before the VLBI run started (for details see Doeleman, this conference, and Greve et al. 2002).
The data were recorded on thin tapes, except at SEST and Mets\"ahovi, where thick 
tapes were used. In 10--15\,min gaps between consecutive VLBI scans (each of $\sim 7$\,min 
duration), pointing checks, antenna temperature and atmospheric opacity measurements (skydips) were 
performed. At HHT, KP and PV typical opacities ranged between $\tau=0.08 - 0.3$. The flux densities of the
VLBI target sources were calibrated by using the planets as primary calibrators. Where possible, these
flux density measurements were later used to determine elevation-gain curves for the antennas. 

Immediately after the observations, selected recorded VLBI tapes from HHT and KP were shipped to Haystack
for a first fringe verification.  The final correlation of the 147 GHz experiment
was done at the MPIfR in Bonn. The fringe fitting was done in two steps, initially with the
MKIV software (FOURFIT, baseline-based fringe fitting) over the full scan length (440 sec).
The data were then imported into AIPS using the new tasks MK4IN and GLAPP (for details, see Alef et al. this conference).
The final fringe fit was done with shorter solution intervals
(0.5 - 3 min) using the AIPS task FRING (station-based global fringe fitting). 
The amplitude calibration of the fringe fitted data  is based on the system temperature and 
antenna gain and opacity measurements recorded during the observations. 
The data were then exported from AIPS to the DIFMAP-package, where the incoherent averaging 
(10 sec) and imaging was done. At the present stage of the data reduction, uncorrected  antenna pointing errors 
and poorly known Kelvin to Jansky conversion factors limit the overall accuracy of the amplitude calibration
to about 20-30\,\%.

\begin{table}
\caption{Signal to noise ratios of VLBI detections at 147 \,GHz}
\begin{tabular}{l|cc|cc|cc}
Source    &\multicolumn{2}{c|}{HHT - KP}& \multicolumn{2}{c|}{HHT - PV}   &\multicolumn{2}{c}{KP - PV}     \\
          & cov1  & cov2     &   cov1  & cov2   & cov1 & cov2   \\ \hline
0133+476  & --    & 7--10    &   --    & --     &  --  &  --    \\
NRAO150   & 7--10 & 9--14    &   --    & 7      &  --  &  --    \\
0420-014  & 7--11 & 8--13    &   --    & --     &  --  &  --    \\
3C273B    & 8     & 6.4?     &   --    & --     &  --  &  --    \\
3C279     & 22--37& 12--49   &   14--40& 20--75 & 8--18&  7--20 \\
1633+382  & 10--20& 10--22   &   11--17& 22--23 & 9--13& 10--12 \\
3C345     & 7     &          &   6.1?  &        &  --  &        \\
MWC349    & --    &          &   --    &        &  --  &        \\
NRAO530   & --    & 11--19   &         &        &      &        \\
SGR\,A*   & 6.7?  &          &         &        &      &        \\
1921-293  & 9     &          &         &        &      &        \\
3C454.3   & 10--15&          &   --    &        &  --  &        \\
BL\,LAC   & 7     &          &   --    &        &  --  &        \\
2255-282  & --    &12--16    &         &        &      &        \\
\end{tabular}

~~\\
cov1, cov2:  coverage on day 109/110 \& day 110/111\\
--: no detection; no entry: not observed\\
\end{table}

In Table 2 we summarize for all observed sources the signal-noise-ratios
of the  VLBI detections obtained so far on the individual interferometer baselines\footnote
{Mets\"ahovi showed SNR $\leq 7$ fringes only with PV due to a problem in VLBI recording. Fringes to
SEST are not yet found}.
The SNR's correspond to incoherent integration over the full scan lengths.
We expect some more detections
after the correlation will be completed and after a fringe search with more
restricted search windows in delay and rate. We note that we have detected 3 sources (NRAO150, 3C279, 1633+382)
on the transatlantic baselines from Arizona (HHT, KP) to Pico Veleta at a fringe spacing
of 4.2G$\lambda$ (corresponding to $\sim 24$\,$\mu$as resolution). This is, as far
as we know, the highest angular resolution, ever achieved in astronomical radio 
interferometry. The SNR of the detections
on the HHT -- PV baseline are higher (SNR $\leq 75$) than on the KP -- PV
baseline (SNR $\leq 20$), indicating a very good sensitivity and phase
stability of the HHT, which was never used in VLBI before. In Figure 1 we plot
the loss of the correlated visibility amplitude as function of integration time
(coherence plot). The atmospheric phase fluctuations limit the integration
time to about 10\,sec, but occasionally allows also much longer integrations.

\begin{figure}
\centering
\includegraphics[angle=-90,width=6cm]{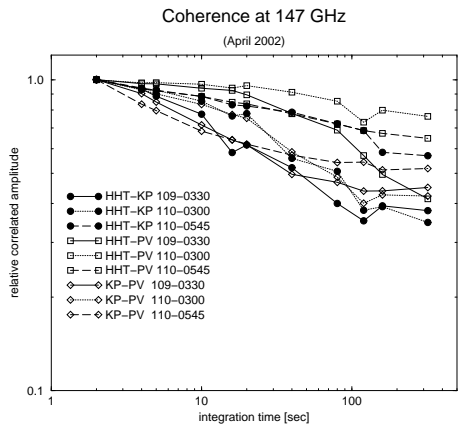}
\caption{Normalized correlated visibility amplitude plotted versus coherent integration time
for the 3 different baselines and for 3 different times (dd-hhmm). The coherence varies
with time and depends mainly on atmospherical conditions. 
Typically, a 10--20\,\% amplitude loss is reached after 10 seconds.}
\end{figure}

\begin{figure}
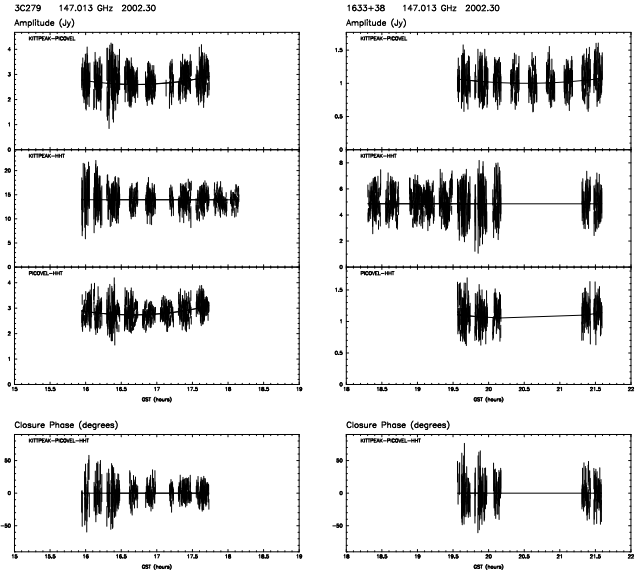

\parbox[t]{4.3cm}{
\includegraphics[angle=0,width=4.2cm, bb= 73 100 587 763, clip]{tkri_fig2.ps}
\includegraphics[angle=0,width=4.2cm, bb= 73 490 587 750, clip]{tkri_fig3.ps}
} 
\parbox[t]{4.3cm}{
\includegraphics[angle=0,width=4.2cm, bb= 73 100 587 763, clip]{tkri_fig4.ps}
\includegraphics[angle=0,width=4.2cm, bb= 73 490 587 750, clip]{tkri_fig5.ps}
}

\caption{Visibility amplitudes and closure phases for 3C\,279 (left) and 1633+382 (right) at 147 GHz for the
station triangle HHT -- KP -- PV. The solid line is a fit of a circular Gaussian model with the flux and 
size given in Table 3. For a few VLBI scans additional corrections of order 30--50\,\% were necessary,
to remove obvious pointing errors.}
\end{figure}

\begin{figure}
\centering
\includegraphics[angle=-90,width=6cm]{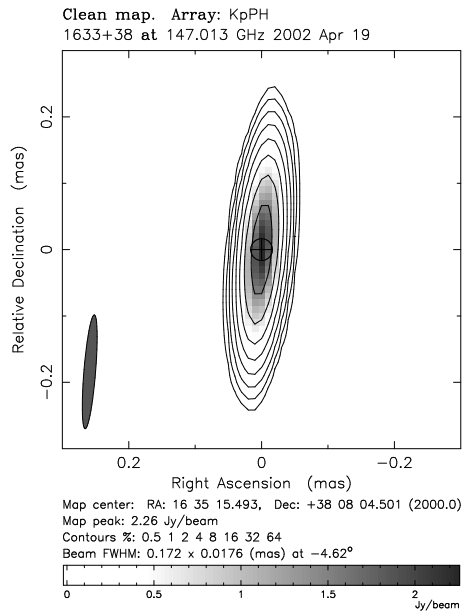}
\caption{Gaussian model of 1633+382 at 147 GHz. Data from HHT -- KP -- PV are shown in Figure 2.
The elongated observing beam reflects the mostly east-west orientation of the
VLB-interferometer. The minor axis of the observing beam is only 18\,$\mu$as.}
\vspace{-0.5cm}
\end{figure}

So far we analyzed only the data for the
two sources, which are detected with SNR $> 7$ on all of the 3 baselines (3C\,279, 1633+382) and for which closure phases could
be measured (see Fig. 2). For both sources the closure phases are consistent with zero, indicating
either a point-like, or a symmetrical source structure (within the limitations given by the uv-coverage).  
In the present preliminary calibration, residual variations of the visibility amplitudes are still relatively
large (of order of 30\,\%). We therefore adopted the simplest approach and fitted just one single circular
Gaussian component to the data. In Figure 3 we show an example of such a model for 1633+382. The
map is convolved with an elliptical restoring beam derived from the uniformly weighted data with the DIFMAP software.  
In Table 3 we show the parameters from the Gaussian fits and the derived brightness
temperatures and linear sizes for both quasars. 

The estimate of the source sizes depends critically on the gain calibration of the 30\,m antenna
on Pico Veleta (accurate within $\sim 20$\,\%), 
to which the long uv-spacings are formed. Larger correlated flux densities on baselines with
this station would yield smaller source sizes. At the moment, it is therefore not completely clear, 
if the two sources shown in Table 3 are in fact marginally resolved along the minor axis of the beam,
or if they are still unresolved. (The  
sizes from the Gaussian fits are about a factor of 1.9 larger than the minor axis of the beam and a factor
of $1.3$ larger than $0.5 (u_{max}^2 + v_{max}^2)^{-0.5}$ ). If the sources were unresolved,
the size estimates shown in Table 3 must be regarded as upper limits to the true source size.
With the usual assumption of the brightness temperature being limited by the inverse Compton effect
($T_B \leq 10^{12}$\,K), an interesting {\it lower} limit to the source size of 
$\geq 17$\,$\mu$as for 1633+382 and $\geq 27$\,$\mu$as for 3C\,279 is obtained 
(from $\theta_{\rm [mas]} \geq [1.22 S_{\rm [Jy]}/\nu_{\rm [GHz]}^2]^{0.5}$). These lower
limits indicate, that at least for 3C279, for which it is larger than the beam size, 
2\,mm-VLBI starts to measure the true spatial extent of this quasar nucleus and this despite of its
large cosmological distance of $z=0.536$ !

\begin{table}
\begin{tabular}{l|c|r|r|c|c|c}
Source  &  z    &$\rm S_{tot}$&  S    &   $\theta$&$T_B$             &    size   \\
        &       &   [Jy]      & [Jy]  & [$\mu$as] & [K]              &    [pc]        \\ \hline
3C279   &0.536  &   21.1      &14.0   &     34    &$7 \cdot 10^{11}$ &   $\simeq 0.20$  \\
1633+382&1.814  &    7.3      & 4.9   &     33    &$3 \cdot 10^{11}$ &   $\leq 0.24$  \\
\end{tabular}
\caption{Results from the transatlantic VLBI detection for 3C279 and 1633+382. Successive
columns give source name, redshift, total flux density, flux density and size ($FWHM$) of the compact
VLBI component from the Gaussian model, its brightness temperature and its linear size (assuming
$H_0=65$ km s$^{-1}$ Mpc$^{-1}$, $q_0 = 0.3$).}
\vspace{-0.7cm}
\end{table}

\section{Outlook}

It is obvious, that with better uv-coverage and more antennas participating in future VLBI at $\leq 2$\,mm,
the calibration uncertainties can be removed and `true maps' rather than simple Gaussian model fits
can be made. From our experience with global 3\,mm-VLBI, we like to point out that many of the radio
sources observed so far, show a considerable amount of sub-structure on the mas- to sub-mas scale
(c.f. Krichbaum et al. 1999).
Imaging of such complex brightness distributions requires a very regular and dense uv-coverage
and at least $\sim 8-10$ VLBI antennas (the more the better). Present day mm-VLBI suffers severely from the
lack of short uv-spacings. As a consequence of this, it is presently not possible to reliably image 
about $30-50$\,\% of the total source flux. The relatively high surface brightness of the sources on 10--500\,km
long baselines, thus  gives room also for less sensitive and smaller antennas to play a significant role
in future high resolution imaging. This experiment has demonstrated that the combination of two
closely spaced smaller antennas (HHT-10m \& KP-12m) with a more distant large antenna (PV-30m), resulted in 
promising new science.

The addition of more collecting area by adding sensitive antennas designed for
millimeter and sub-millimeter research will always be most important for mm-VLBI. This includes existing antennas, which are
not yet participating in mm-VLBI (eg. JCMT, SMA, NRO), but also new antennas like eg. the 50\,m LMT in Mexico
and the ALMA prototype antenna APEX. Owing to their outstanding sensitivity,
phased interferometers will play a particular important role in future mm-VLBI. For the near future, the participation
of the IRAM interferometer at Plateau de Bure (France) is planned (first at 3 \& 1\,mm, later
also at 2\,mm). This will increase the present sensitivity by a factor of $\sim 2-3$ to the $\sim 0.1$\,Jy level. 
In the more distant future CARMA - the merger of the BIMA and OVRO interferometers - and ALMA
should lower the detection threshold to 1--10\,mJy and by this tremendously improve the 
observational possibilities (c.f. Krichbaum et al., 1996). In parallel, higher data recording rates and larger
observing bandwidths (GBit/sec using the future MKV system) and the possibility of atmospheric
phase corrections (via water vapor radiometry and/or phase referencing) will also help to further
improve the sensitivity. 

All this will facilitate the imaging of Quasars, nearby Radio-Galaxies and of the Galactic Center
source with the fascinating angular resolution of only $\sim 10-20$\,$\mu$arcseconds ! 
In nearby galaxies, structures of a few light days in size could then be seen. If the
Galactic Center source Sgr\,A* would be detected with mm-VLBI at $\geq 4$\,G$\lambda$ resolution, the 
direct imaging of a region as small as $\sim 2-4$ Schwarzschild radii would become possible.
It is therefore not completely unrealistic that in 10--20\,yrs, general relativistic
effects caused by the distortion of the space-time in nearby supermassive Black Holes 
would become directly observable.

\begin{acknowledgements}
We thank all the personnel from the observatories who participated in this experiment for their
efforts and enthusiasm, without which this success would not have been possible.
\end{acknowledgements}

\end{document}